\documentclass[aps,prd,twocolumn,superscriptaddress,showpacs,amsmath,amssymb,amsthm,nofootinbib, preprintnumbers]{revtex4-2}

 \usepackage{amsmath}

\usepackage{graphicx}

\usepackage{epstopdf}

\usepackage{float}

\usepackage{hyperref}

\usepackage{color}

\usepackage[T1]{fontenc}

\usepackage[utf8]{inputenc}

\usepackage[toc,page]{appendix}

\usepackage[usenames,dvipsnames]{xcolor}

\usepackage[normalem]{ulem}

\usepackage{lipsum, babel}

\newcommand{\be}{\begin{equation}}

\newcommand{\ee}{\end{equation}}

\newcommand{\ba}{\begin{eqnarray}}

\newcommand{\ea}{\end{eqnarray}}

\newcommand{\beq}{\begin{equation}}

\newcommand{\eeq}{\end{equation}}

\newcommand{\beqa}{\begin{eqnarray}}

\newcommand{\eeqa}{\end{eqnarray}}

\begin{document}

\title{Holographic dual of extended black hole thermodynamics}

\author{Moaathe Belhaj Ahmed}

\affiliation{Department of Physics and Astronomy, University of Waterloo, Waterloo, Ontario, N2L 3G1, Canada}

\author{Wan Cong}

\email{wan.cong@univie.ac.at}

\affiliation{University of Vienna, Faculty of Physics, Boltzmanngasse 5, A 1090 Vienna, Austria}

\author{David Kubiz\v n\'ak}

\email{david.kubiznak@matfyz.cuni.cz}

\affiliation{Institute of Theoretical Physics, Faculty of Mathematics and Physics,
Charles University, Prague, V Hole{\v s}ovi{\v c}k{\' a}ch 2, 180 00 Prague 8, Czech Republic}

\author{Robert B. Mann}

\email{rbmann@uwaterloo.ca}

\affiliation{Department of Physics and Astronomy, University of Waterloo, Waterloo, Ontario, N2L 3G1, Canada}

\author{Manus R. Visser}

\email{mv551@cam.ac.uk}

\affiliation{Department of Applied Mathematics and Theoretical Physics,\\ University of Cambridge,
Wilberforce Road, Cambridge CB3 0WA, United Kingdom}

\date{\today}

\begin{abstract}
By respecting the conformal symmetry of the dual CFT, and   treating the conformal factor of the AdS boundary as a  thermodynamic parameter,
we formulate the holographic   first law   that is  exactly dual to the first law of extended black hole thermodynamics with variable cosmological constant but fixed Newton's constant. 

\end{abstract}

\maketitle

{\em Extended black hole thermodynamics}, also known as {\em black hole chemistry} \cite{Kubiznak:2014zwa,Kubiznak:2016qmn}, is one of the major developments in classical black hole thermodynamics  in recent years. The idea stems from reconsidering the thermodynamics of asymptotically  Anti-de Sitter (AdS) black holes in the context of a variable cosmological constant $\Lambda$ \cite{Teitelboim:1985dp,Creighton:1995au}, which is treated as a thermodynamic pressure according to the following (perfect fluid) prescription: 
\be
P=-\frac{\Lambda}{8\pi G_N}=\frac{(d-1)(d-2)}{16\pi G_N L^2}\,, 
\ee
where $G_N$ is Newton's   constant,  $L$ is the AdS radius, and $d$ stands for the number of (bulk) spacetime dimensions. 

This identification allows one to define the black hole volume $V$ and introduces the standard pressure-volume term into black hole thermodynamics. Namely, we now have the extended first law, together with the corresponding generalized  Smarr relation \cite{Kastor:2009wy}:
\ba
\delta M&=&T\delta S+V\delta P+\Phi \delta Q+\Omega \delta J\,, \label{bulk}\\
M&=&\frac{d-2}{d-3}(TS+\Omega J)+\Phi Q-\frac{2}{d-3}PV \label{Smarrbulk}
\,, 
\ea
with the two being related by a dimensional scaling argument (resulting in $d$-dependent factors in the Smarr relation). 
The key result to emerge from the black hole chemistry approach    is the  realization that AdS black holes exhibit phase transitions that are fully analogous to those of ordinary thermodynamic systems. In particular, one observes phase transitions \`{a} la Van der Waals  \cite{Chamblin:1999tk,Kubiznak:2012wp}, reentrant phase transitions \cite{Altamirano:2013ane}, isolated critical points \cite{Dolan:2014vba,Ahmed:2022kyv},  superfluid like behavior \cite{Kubiznak:2016qmn}, and multicritical points \cite{Tavakoli:2022kmo}. 
Very recently a mechanism for the higher-dimensional origin of a dynamical cosmological constant was  proposed  \cite{Frassino:2022zaz}.

However, the {\em holographic interpretation} of extended thermodynamics remained unclear for many years. The first attempts \cite{Johnson:2014yja,Dolan:2014cja,Kastor:2014dra,Zhang:2014uoa, Dolan:2016jjc} suggested that, according to the Anti-de Sitter/Conformal Field Theory (AdS/CFT) correspondence \cite{Maldacena:1997re}, the $V\delta P$ term should be related  to a
$
\mu \delta C 
$
term in the   dual CFT, where $C$ is the central charge and $\mu$   the  thermodynamically conjugate chemical potential.  For     holographic CFTs dual to Einstein gravity the dictionary for the central charge is
 \be\label{C}
C\propto\frac{L^{d-2}}{ G_N}\,, 
\ee
where the proportionality constant depends on the normalization of $C$, which is irrelevant for the discussion below.  
The  situation is not that simple, however,  as it is also standard to identify the curvature radius of the spatial geometry on which the CFT is formulated with the AdS radius $L$   \cite{Karch:2015rpa}.
Namely, the boundary metric of the dual CFT is obtained by the conformal completion of the bulk AdS spacetime and reads \cite{Gubser:1998bc,Witten:1998qj}
\be
ds^2= \omega^2 \Bigl(-dt^2+L^2d\Omega_{k,d-2}^2\Bigr)\,,
\ee
where   $\omega$ is an `arbitrary'  dimensionless  conformal factor, a function of boundary coordinates, that reflects the conformal symmetry of the boundary theory.   For $k=1$   $d \Omega_{k,d-2}^2 \!$ is the metric on a unit ($d-2$)-dimensional sphere, for $k=0$ it is the dimensionless metric $\frac{1}{L^2} \sum_i dx_i^2  $  on the plane, and for $k=-1$ it is the unit metric on  hyperbolic space   $  d u^2 + \sinh^2 (u) d \Omega_{k=1,d-3}^2$.   

The standard choice is to set $\omega=1$, in which case the CFT volume ${\cal V}$ is proportional to $L^{d-2}$. Consequently a variation of the cosmological constant in the bulk induces also 
a variation of the CFT volume ${\cal V}$.  Hence a pressure-volume work term, $-p\delta {\cal V}$,   should be present on the CFT side.  This implies that  either (i)  the corresponding CFT first law is {\em degenerate} as the $\mu \delta C$ and $-p \delta {\cal V}$ terms are not truly independent (leaving the CFT interpretation of   black hole chemistry a bit obscure), or (ii)    apart from varying the cosmological constant  we also have to vary Newton's gravitational constant $G_N$ so that the variations of $\cal V$ and $C$ are independent~\cite{Karch:2015rpa}.   

Alternatively, in the spirit of the second option,  the authors of  \cite{Zeyuan:2021uol} have proposed the so-called \emph{restricted phase space} (RPS) formalism, where the CFT volume ${\cal V}\propto L^{d-2}$ is kept fixed. This   leaves only the $\mu\delta C$ term  on the CFT side coming from a variable  $G_N$  in the bulk. The resultant holographic thermodynamics thus has nothing to do with the original black hole chemistry.

In this note we  generalize the approach developed in \cite{Visser:2021eqk} to find the holographic first law that is dual to~\eqref{bulk} while avoiding both of the above mentioned problems.    
 Namely, in order to capture the above rescaling freedom of the CFT, while in the setting of equilibrium thermodynamics, in what follows we shall treat $\omega$ as a (dimensionless) thermodynamic parameter   (similar to the horizon radius or AdS radius), rather than a function of the boundary coordinates. This will make the volume and central charge independent variables.  
 
    This is not without precedent.
For the $k=0$ planar AdS black brane case, variations of   volume $\mathcal{V}$ and central charge $C$ are  clearly   independent; varying the former corresponds to changing the number of points in the system, whereas varying the latter corresponds to varying the number of degrees of freedom at each point. Since the planar case can be reached as a limit of the $k=1$ spherical case, it is reasonable to expect this independence to extend to non-planar cases. 

Consequently, rather than using the standard choice  $\omega=1$, we regard $\omega$ as another variable. This effectively amounts to changing the CFT volume, which  is now proportional to
\be \label{volume}
{\cal V} \propto (\omega L)^{d-2}\,. 
\ee
In \cite{Visser:2021eqk,Cong:2021jgb} we considered the choice     $\omega=R/L$ with $R$  being  a constant  boundary curvature radius, but here we allow the conformal factor to be  a generic   parameter  which need not depend on $L$. For the Einstein-Maxwell   Lagrangian density $\mathcal L = \frac{1}{16 \pi G_N} (R -2 \Lambda) - \frac{1}{4}F^2$  this results  in the following generalized dictionary between the bulk (without tildes) and dual CFT (with tildes) thermodynamic quantities:
\ba \label{extendeddictionary}
\tilde S&=& S=\frac{A}{4 G_N}\,,\quad 
\tilde E=\frac{M}{\omega}\,,\quad \tilde T=\frac{T}{\omega}\,,\quad \tilde\Omega=\frac{\Omega}{\omega}\,,\nonumber\\
\tilde J &=& J, \quad \tilde \Phi=\frac{\Phi\sqrt{G_N}}{\omega L}\,,\quad \tilde Q=\frac{QL}{\sqrt{G_N}}\,.
\ea
If we now allow the bulk curvature radius $L$ to vary, while  \textit{holding $G_N$ fixed}, the variation of the central charge~$C$, \eqref{C}, is then purely induced by  variations of~$L$.  

Analogously to the calculation in \cite{Visser:2021eqk} (see also \cite{Cong:2021fnf,Cong:2021jgb}), it is   straightforward to show  using \eqref{Smarrbulk}  that the extended first law \eqref{bulk}   can be rewritten as follows:
\begin{align}\label{Eq8}
&\delta \Bigl(\frac{M}{\omega}\Bigr)= \frac{T}{\omega}\delta \Bigl(\frac{A}{4 G_N}\Bigr)+\frac{\Omega}{\omega}\delta J+\frac{\Phi\sqrt{G_N}}{\omega L}\delta \Bigl(\frac{QL}{\sqrt{G_N}}\Bigl)\nonumber\\
&\qquad +   \Bigl(\frac{M}{\omega}-\frac{TS}{\omega}-\frac{\Omega J }{\omega}-\frac{\Phi Q}{\omega}\Bigr) \!\frac{\delta(L^{d-2}/G_N)}{L^{d-2}/G_N}\nonumber\\
&\qquad-\frac{M}{\omega(d-2)}\frac{\delta (\omega L)^{d-2}}{(\omega L)^{d-2}},\quad \  
\end{align}
or simply  as   
\be
\delta \tilde E=\tilde T\delta S+\tilde \Omega \delta J+\tilde \Phi \delta \tilde Q+\mu \delta C-p\delta {\cal V}\,,\label{j1}
\ee
where, using \eqref{C}, \eqref{volume} and \eqref{extendeddictionary}, 
\ba
\mu&=&\frac{1}{C}( \tilde E - \tilde TS-\tilde \Omega J-\tilde \Phi \tilde Q)\,, \label{j2}\\
p&=& \frac{\tilde E}{(d-2){\cal V}}\,.\label{j3}
\ea 
The CFT first law \eqref{j1} is no longer degenerate, as both $\cal V$ and $C$ can now be independently varied. Together with relations \eqref{j2} and \eqref{j3}, it is  
exactly dual to the first law of  extended black hole thermodynamics. 

As is obvious from the derivation, the variation of $\Lambda$  does not only enter in the variation of the central charge, but it also appears in the dictionary for the  spatial volume  and electric charge. The variation of $\Lambda$ (the $V\delta P$ term in \eqref{bulk}) has thus been split into several pieces and is related  to the variation of the volume, electric charge and central charge of the CFT.

Eq.~\eqref{j3} is the equation of state  for conformal theories, which is derivable from the scaling symmetry of the CFT. 
Moreover,  \eqref{j2} is the Euler relation for holographic CFTs,  which can be derived on the CFT side from the proportionality of the thermodynamic quantities with the central charge, $\tilde E, \tilde   S, \tilde  J, \tilde Q \!\propto \!C$,  which occurs   in the deconfined   phase (that is dual to an AdS black hole geometry). We note the absence of a $-p\mathcal V$ term in the Euler relation, which reflects the fact that  the internal energy is not an extensive variable on compact spaces at finite temperature in the deconfined phase. This is not an issue, as claimed in \cite{Zeyuan:2021uol}, but rather a feature of holographic CFTs. In the high-temperature or large-volume regime, i.e. $\omega L \tilde T  \gg 1$, the $\mu C$ term becomes equal to $ - p \mathcal V$, and hence the energy becomes extensive.   
As explained in  \cite{Visser:2021eqk,Karch:2015rpa}, the Euler relation \eqref{j2}  is  dual to the  Smarr formula~\eqref{Smarrbulk} for AdS black holes. The latter relation contains dimension dependent factors, whereas the former does not.  We can understand this by expressing the   $PV$ term in the Smarr formula in terms of a partial derivative of the CFT energy
\beq
\label{PVterm}
- 2 P V = L \left ( \frac{\partial M}{\partial L}\right)_{A,J,Q,G_N}\!\!\!\! = L \omega \left (\frac{\partial \tilde E}{\partial L}\right)_{A,J,Q,G_N} \!\!\!\!\,.
\eeq   
The boundary energy depends on the bulk quantities as $\tilde E = \tilde E (S(A,G_N),J,\tilde Q(Q,L,G_N),C(L,G_N),V(L,\omega))$. 
Hence, the dictionary \eqref{C}, \eqref{volume} and \eqref{extendeddictionary} implies that  
\begin{align}
\label{partial}
&\left ( \frac{\partial \tilde E}{\partial L}\right)_{A,J,Q,G_N} \!\!\!\!\!\!=\frac{1}{L} ( \tilde \Phi \tilde Q + (d-2)\mu C - (d-2) p {\cal V})  \\
&\qquad \qquad \,\, =\frac{1}{L} ( (d-3) (\tilde E - \tilde \Phi \tilde Q) - (d-2) (\tilde \Omega J + \tilde T S))\,, \nonumber
 \end{align}
 where  we   inserted the Euler relation and the equation of state to obtain the last equality.  By combining \eqref{PVterm} and \eqref{partial}, and using the holographic dictionary  we recover the Smarr relation.

Since the Euler relation and equation of state   follow from the scaling of the thermodynamic quantities with $C$ and with $\cal V$, respectively, one can eliminate some of the terms in the first law \eqref{j1} by rescaling the CFT quantities. In particular, using \eqref{j3} and rescaling  some of  the CFT quantities by   $\omega L$ we can eliminate the $-p\delta V$ term, to obtain  the laws  
\ba
\delta \hat E&=&\hat T\delta S+\hat \Omega \delta J+\hat \Phi \delta \tilde Q+\hat \mu \delta C\,,\label{j4} \\
\hat E&=&\hat T S+\hat \Omega J+\hat \Phi \tilde Q+\hat \mu C\,,  \label{j5}
\ea
for the rescaled (dimensionless)  quantities:
\ba
\label{re1}
\hat E&=&\omega L \tilde E \,,\quad \hat T=\omega L \tilde T\,,\quad \hat \Omega=\omega L \tilde \Omega \,,\nonumber\\
 \hat \Phi&=&\omega L \tilde \Phi\,, \quad \hat \mu=\omega L \mu\,. 
\label{eq14}
\ea
The advantage of   \eqref{eq14} is that all thermodynamic quantities are now scale invariant, and so the thermal description respects the symmetries of the CFT. 

While the  laws in \eqref{j4} and \eqref{j5} are formally the same as those of the recently proposed RPS \cite{Zeyuan:2021uol}, their physical interpretation is different. In our case the CFT lives on a geometry with an arbitrary curvature radius~$\omega L$, distinct  from the AdS radius $L$.  At the same time  we allow   the central charge $C$ to vary. Contrary to the RPS approach, this variation is induced by a variable cosmological constant in the bulk, rather than a variable Newton's constant.

 Perhaps more interesting is to employ the Euler relation \eqref{j2} to eliminate the $\mu \delta C$ term from \eqref{j1}, yielding
\ba
\delta \bar{E}&=&\tilde {T}\delta \bar{S}+\tilde {\Omega} \delta \bar{J}+\tilde {\Phi} \delta \bar{Q}-\bar{p} \delta {\cal V}\,,\label{j8}\\
\bar E&=& (d-2) \bar p {\cal V}\,, 
\ea
with the rescaled quantities: 
\be 
\label{re2}
\bar{E}=\frac{\tilde E}{C}\,,\quad \!
\bar{S}=\frac{S}{C}\,,\quad \!
\bar{J}=\frac{J}{C}\,,\quad \!\bar{Q}=\frac{\tilde Q}{C}\,,\quad \bar{p}=\frac{p}{C}\,.
\ee
 These quantities are no longer proportional to $C$, i.e., they are $\mathcal O(C^0)$. The advantage of these laws is that all thermodynamic quantities keep their correct dimensionality. Moreover, for fixed $C$ one recovers the `standard' thermodynamic first law, with $\bar{E}$ interpreted as internal energy. Of course, the rescalings in \eqref{re1} and \eqref{re2} can be combined together, to obtain a dimensionless CFT law for the rescaled quantities   without $- p \delta \cal V$ and $\mu \delta C $ terms. However, both the central charge $C$ and the CFT volume ${\cal V}$ can remain dynamical quantities in this first law.

To summarize, we have established an exact duality between the extended black hole thermodynamics and the CFT description.  Whereas the bulk first  law \eqref{bulk} has one extra ($V\delta P$) term and is accompanied by a single Smarr relation \eqref{Smarrbulk}, to reflect the scaling symmetry of the dual CFT, the corresponding CFT first law \eqref{j1} has two extra terms and is accompanied by  two relations:  the Euler equation \eqref{j2} and the equation of state \eqref{j3}.
For this reason, the variation of $\Lambda$ in the bulk corresponds to both changing the CFT central charge $C$ and the CFT volume ${\cal V}$.

Furthermore, the corresponding $\mu\delta C$ and $-p \delta{\cal V}$ terms can be further eliminated by using \eqref{j2} and \eqref{j3}. In this way one formally recovers the laws of the RPS, or the striking~\eqref{j8} -- but we emphasize that our approach  differs from \cite{Zeyuan:2021uol} in that we keep $G_N$ fixed but allow the AdS radius $L$ and conformal factor $\omega$ to independently vary.   Finally, it would be interesting to generalize the holographic dual of extended black hole thermodynamics to other geometries, e.g., de Sitter spacetime, and other gravitational theories, e.g., higher curvature gravity.

\subsection*{Acknowledgements}

We would like to thank Roberto Emparan and David  Mateos for useful discussions. This work was supported in part by the Natural Sciences and Engineering Research Council of Canada. D.K. is grateful for support from GA{\v C}R 23-07457S grant  of the Czech Science Foundation. M.R.V. is supported by  SNF Postdoc Mobility grant P500PT-206877 ``Semi-classical thermodynamics of black holes and the information paradox''.

\bibliography{references}

\end{document}